# Squeezing Microwaves by Magnetostriction

Jie Li[1, 2,*], Yi-Pu Wang[1], J. Q. You[1] and Shi-Yao Zhu[1]

[1]*Interdisciplinary Center of Quantum Information, Zhejiang Province Key Laboratory of Quantum Technology and Device, and State Key Laboratory of Modern Optical Instrumentation, Department of Physics, Zhejiang University, Hangzhou 310027, China*
[2]*Kavli Institute of Nanoscience, Department of Quantum Nanoscience, Delft University of Technology, Delft 2628CJ, The Netherlands*
*Corresponding author. E-mail: jieli007@zju.edu.cn

**ABSTRACT**
Squeezed light finds many important applications in quantum information science and quantum metrology, and has been produced in a variety of physical systems involving optical nonlinear processes. Here, we show how a nonlinear magnetostrictive interaction in a ferrimagnet in cavity magnomechanics can be used to reduce quantum noise of the electromagnetic field. We show optimal parameter regimes where a substantial and stationary squeezing of the microwave output field can be achieved. The scheme can be realized within the reach of current technology in cavity electromagnonics and magnomechanics. Our work provides a new and practicable approach for producing squeezed vacuum states of electromagnetic fields, and may find promising applications in quantum information processing and quantum metrology.

## INTRODUCTION

Squeezed states of light [1], with the noise at certain phases below vacuum fluctuation, are typically produced by using certain optical nonlinear interactions [2]. They were first produced by four-wave mixing in sodium atoms [3], shortly after by employing optical fibers [4], and then by optical parametric amplification in nonlinear crystals [5]. Squeezed light can also be generated from a semiconductor laser [6], a single atom in an optical cavity [7], a single semiconductor quantum dot [8], and optomechanical systems [9–13]. Recently, a substantial squeezing of 15 dB has been produced in a nonlinear crystal [14]. Squeezed light finds a wide range of important applications: It can be used to improve the sensitivity of interferometers for gravitational-wave detection [15], to produce an Einstein-Podolsky-Rosen entangled source used for, e.g., quantum teleportation [16], to enhance sensitivity in biological measurements [17], and to calibrate the quantum efficiency of photoelectric detection [14], among many others. In the microwave domain, squeezed states are mainly produced by degenerate parametric down-conversion in a Josephson parametric amplifier (JPA) [18–22] utilizing the nonlinearity of the Josephson junction. Impressively, a 10-dB squeezed microwave field has been generated [20].

In this article, we introduce a completely new approach, based on a recently demonstrated cavity magnomechanical system [23–25], for producing squeezed vacuum states of microwave fields. We show how the nonlinear magnetostrictive interaction in cavity magnomechanics can be used to reduce the quantum noise of the electromagnetic field. The system consists of a magnon mode (e.g., the Kittel mode [26]) in a yttrium-iron-garnet (YIG) ferrimagnet, which simultaneously couples to a microwave cavity mode via magnetic dipole interaction [27–29] and to a deformation phonon mode of the ferrimagnet through the magnetostrictive interaction [30]. The magnetostrictive force couples the magnon excitations inside the YIG ferrimagnet to the deformation of its geometry structure. It is a radiation pressure-like dispersive interaction [31], which provides necessary nonlinearity for generating a squeezed spin wave. Specifically, a deformation displacement of the YIG ferrimagnet is caused by the magnetostrictive interaction, proportional to the magnon excitation number, which in turn modulates the phase of magnon mode, giving rise to a correlation between the amplitude and the phase of the magnon mode. This correlation results in a quadrature squeezing of the magnon mode. The mechanism is akin to the ponderomotive squeezing of light induced by radiation pressure in optomechanics [9–12, 32, 33], given the radiation pressure-like magnetostrictive interaction in magnomechanics. The magnetic dipole interaction enables a state-swap (beamsplitter) interaction between the magnon mode and the microwave cavity field. Therefore, the squeezing of the magnon mode is transferred to the microwave cavity field, and the squeezing can be accessed in the cavity output field via a homodyne detection. We show optimal parameter regimes of the system where a substantial squeezing in the output field can be obtained. The squeezing can be created at a temperature much higher than the operation temperatures of a JPA [19–22]. This is an advantage of our approach for producing microwave squeezing.

**THE CAVITY MAGNOMECHANICAL SYSTEM**

We consider a general model of a cavity magnomechanical system, which consists of a microwave cavity mode, a magnon mode, and a phonon mode, as depicted in Fig. 1(a) (see the online supplementary material for a specific experimental setup). The magnons, as quanta of spin waves, describe collective spin excitations in a ferrimagnet, e.g., YIG. The ferrimagnet can be a YIG sphere [24, 25] or a YIG film [34–37]. The magnon mode couples to a microwave cavity mode via the magnetic dipole interaction, and to a lower-frequency phonon mode (an elastic deformation mode due to lattice vibrations) of the ferrimagnet by a magnetostrictive force [30], which couples the magnon excitations to the deformation displacement of the ferrimagnet in a dispersive manner. We note that although a higher-frequency phonon mode is considered in Refs. [34–37] (as they used a thin film with thickness of a few μm), a thicker YIG film should be employed such that the phonon frequency becomes much lower than the magnon frequency, and then the dominant magnomechanical interaction remains dispersive. This magnon-phonon dispersive interaction is essential as it provides the nonlinearity required to create squeezed states in the system. The size of the ferrimagnet is assumed to be much smaller than the wavelength of the microwave cavity field, of which the frequency is typically around 10 GHz [27–29]. Thus, their radiation pressure interaction can be fully neglected. The Hamiltonian of this tripartite system is

$$H/\hbar = \omega_a a^\dagger a + \omega_m m^\dagger m + \frac{\omega_b}{2}(q^2 + p^2) + G_0 m^\dagger m q$$
$$+ g(am^\dagger + a^\dagger m) + i\Omega(m^\dagger e^{-i\omega_d t} - m e^{i\omega_d t}), \qquad (1)$$

where $a$ and $m$ ($a^\dagger$ and $m^\dagger$) are the annihilation (creation) operators of the cavity and magnon mode, respectively, satisfying $[C, C^\dagger] = 1$ ($C = a, m$), $q$ and $p$ are the dimensionless position and momentum quadratures of the phonon mode, thus satisfying $[q, p] = i$, and $\omega_j$ ($j = a, m, b$) are the resonance frequencies of the cavity, magnon and phonon mode, respectively. While the cavity resonance is determined by the configuration of the cavity, the magnon mode frequency can be tuned by adjusting the external bias magnetic field $H_0$ via $\omega_m = \gamma_0(H_0 - H_d)$, where $\gamma_0/2\pi = 28$ GHz/T is the gyromagnetic ratio of YIG and $H_d$ is the demagnetization field. The cavity-magnon coupling rate $g \propto \sqrt{N}$, with $N$ being the number of spins, and $N = 4\pi\rho R^3/3$ for a sphere, where $\rho = 4.22 \times 10^{27}$ m$^{-3}$ is the spin density of YIG and $R$ the radius of the sphere. Owing to the high spin density of YIG, the strong coupling $g > \kappa_a, \kappa_m$ can be easily achieved [27–29], where $\kappa_a$ ($\kappa_m$) is the linewidth (FWHM) of the cavity (magnon) mode. However, typically $g \ll \omega_a, \omega_m$, which allows to approximate the intrinsic cavity-magnon coupling $g(a + a^\dagger)(m + m^\dagger)$ to be $g(am^\dagger + a^\dagger m)$, responsible for the cavity-magnon state-swap interaction. The bare magnomechanical coupling $G_0$ is typically weak for a large-sized YIG sphere with diameter of a few hundred μm [24, 25], but it can be significantly enhanced by directly driving the magnon mode with a strong microwave field, e.g., by using a small microwave loop antenna [38–41]. There are other efficient ways to improve the magnomechanical coupling $G_0$: i) by reducing the sphere's diameter as $G_0 \propto R^{-2}$ [24], ii) by adopting a nonspherical sample, e.g., a YIG film [34–37], which shows a stronger coupling because the thickness of the film can be very small, while it is challenging to produce high-quality crystalline YIG spheres with diameter smaller than 250 μm, iii) by using a ferromagnetic material, e.g., CoFeB [42], which exhibits a large magnetostriction. $\Omega = \frac{\sqrt{5}}{4}\gamma_0\sqrt{N}B$ [31] is the coupling strength between the magnon mode and its driving magnetic field of amplitude $B$ and frequency $\omega_d$. Note that for the magnon mode, we have expressed the collective spin operators in terms of Boson operators via the Holstein-Primakoff transformation [43] under the condition of low-lying excitations, $\langle m^\dagger m \rangle \ll 2Ns$, with $s = \frac{5}{2}$ the spin number of the ground state $Fe^{3+}$ ion in YIG. This approach has been widely adopted in, e.g., Refs. [23, 29, 31, 38, 39].

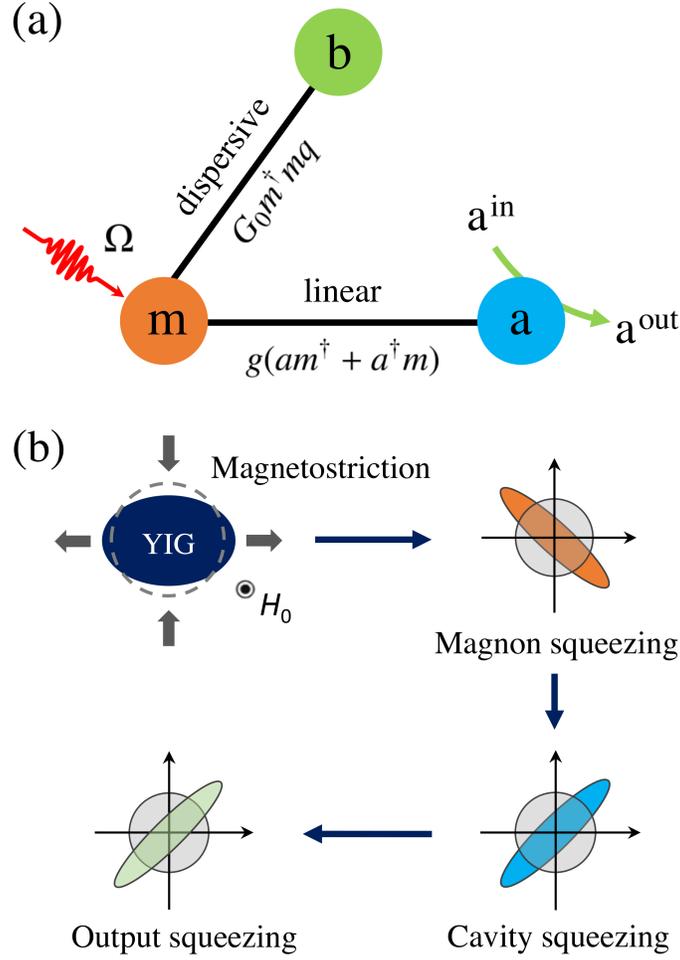

**Figure 1.** (a) Sketch of the interactions in cavity magnomechanics with a dispersive magnomechanical coupling. The magnon mode $m$ couples to the cavity mode $a$ via the magnetic dipole interaction (in a linear form), and to a vibrational phonon mode $b$ of the ferrimagnet by the magnetostrictive interaction (in a nonlinear form). The magnon mode is directly driven by a strong microwave field to enhance the magnomechanical coupling strength. (b) The mechanism of generating squeezed microwave fields. The nonlinear magnetostrictive interaction yields a squeezed magnon mode (spin wave), which leads to a squeezed microwave cavity field via the state-swap interaction, and thus a squeezed cavity output field. The colored ellipses denote single-mode squeezing in terms of fluctuations (shaded) of the quadratures.

By strongly driving the magnon mode and including the dissipation and input noise of each mode, the Hamiltonian (1) leads to the following linearized quantum Langevin equations (QLEs) for the quantum fluctuations in the frame rotating at the drive frequency $\omega_d$ (see the online supplementary material)

$$\begin{aligned}
\delta\dot{q} &= \omega_b \delta p, \\
\delta\dot{p} &= -\omega_b \delta q - \gamma \delta p - G^* \delta m - G \delta m^\dagger + \xi, \\
\delta\dot{m} &= -(\tfrac{\kappa_m}{2} + i\widetilde{\Delta}_m)\delta m - ig\delta a - iG\delta q + \sqrt{\kappa_m}m^{\text{in}}, \\
\delta\dot{a} &= -\left(\tfrac{\kappa_a}{2} + i\Delta_a\right)\delta a - ig\delta m + \sqrt{\kappa_1}a_1^{\text{in}} + \sqrt{\kappa_2}a_2^{\text{in}},
\end{aligned} \quad (2)$$

where $\gamma$ is the mechanical damping rate, $G$ is the enhanced magnomechanical coupling rate, which is complex, and $\widetilde{\Delta}_m$ is the effective magnon-drive detuning, which includes the frequency shift induced by the magnetostrictive interaction (see the online supplementary material for the expressions of $G$ and $\widetilde{\Delta}_m$), and $\Delta_a = \omega_a - \omega_d$. $\xi$ denotes a Brownian stochastic force, which is non-Markovian by nature, but it can be assumed Markovian for a large mechanical quality factor $Q \gg 1$ [44]. In this case, it becomes $\delta$-correlated: $\langle \xi(t)\xi(t') + \xi(t')\xi(t)\rangle/2 \simeq \gamma[2\bar{n}_b(\omega_b) + 1]\delta(t-t')$. $m^{\text{in}}$ and $a_j^{\text{in}}$ ($j=1,2$) are input noise operators for the magnon and cavity mode, respectively. $a_1^{\text{in}}$ is the input noise entering the connector port of the microwave cavity, through which the output field is sent into a vector network analyser, or a homodyne detection scheme [21, 22], and the corresponding external coupling rate is $\kappa_1$. $a_2^{\text{in}}$ is the input noise describing all the other decay channels with a total decay rate $\kappa_2 \equiv \kappa_a - \kappa_1$. $m^{\text{in}}$ and $a_j^{\text{in}}$ are zero-mean and possess the following nonzero correlation functions: $\langle m^{\text{in}}(t)m^{\text{in}\dagger}(t')\rangle = [\bar{n}_m(\omega_m) + 1]\delta(t-t')$, $\langle m^{\text{in}\dagger}(t)m^{\text{in}}(t')\rangle = \bar{n}_m(\omega_m)\delta(t-t')$, and $\langle a^{\text{in}}(t)a^{\text{in}\dagger}(t')\rangle = [\bar{n}_a(\omega_a) + 1]\delta(t-t')$, $\langle a^{\text{in}\dagger}(t)a^{\text{in}}(t')\rangle = \bar{n}_a(\omega_a)\delta(t-t')$, where $\bar{n}_j(\omega_j) = [\exp(\hbar\omega_j/k_BT) - 1]^{-1}$ ($j = b, m, a$) are the mean thermal phonon, magnon, photon number, respectively, at an environmental temperature $T$.

The QLEs (2) can be conveniently solved in the frequency domain by taking the Fourier transform of each equation. The expressions for the quantum fluctuations $\delta Q(\omega)$ ($Q = a, m, q, p$) can be obtained, which take the form of

$$\delta Q(\omega) = \sum_{j=1,2} \left[ Q_{A_j}(\omega) a_j^{in}(\omega) + Q_{B_j}(\omega) a_j^{in\dagger}(-\omega) \right] \\ + Q_C(\omega) m^{in}(\omega) + Q_D(\omega) m^{in\dagger}(-\omega) + Q_E(\omega) \xi(\omega), \quad (3)$$

where $Q_k(\omega)$ ($k = A_j, B_j, C, D, E; j = 1,2$) are frequency-dependent coefficients associated with different input noises.

We are interested in the output field of the microwave cavity and its noise property. We thus calculate the noise spectral density (NSD) of the cavity output field. Its quantum fluctuation $\delta a^{out}(\omega)$ can be obtained by using the standard input-output relation, $\delta a^{out}(\omega) = \sqrt{\kappa_1} \delta a(\omega) - a_1^{in}(\omega)$. Considering the fact that we use a homodyne detection for the output field, where the phase of the local oscillator matters, we define a general quadrature of the output field

$$\delta W^{out}(\omega) = \frac{1}{\sqrt{2}} \left[ \delta a^{out}(\omega) e^{-i\phi} + \delta a^{out\dagger}(-\omega) e^{i\phi} \right], \quad (4)$$

with $\phi$ the phase angle. When $\phi = 0$ ($\frac{\pi}{2}$), $\delta W^{out}(\omega) = \delta X^{out}(\omega)$ ($\delta Y^{out}(\omega)$), corresponding to the amplitude (phase) fluctuation of the output field. The NSD of the general quadrature is defined as

$$S_W^{out}(\omega) = \frac{1}{4\pi} \int_{-\infty}^{+\infty} d\omega' e^{-i(\omega+\omega')t} \times \\ \langle \delta W^{out}(\omega) \delta W^{out}(\omega') + \delta W^{out}(\omega') \delta W^{out}(\omega) \rangle. \quad (5)$$

By using the input noise correlations in the frequency domain, $S_W^{out}(\omega)$ can be obtained, which is, however, too lengthy to be reported. The output field is squeezed if $S_W^{out}(\omega)$ is smaller than that of the vacuum state, i.e. $S_W^{out}(\omega) < \frac{1}{2}$ in our notations.

**RESULTS**

In this section, we show that squeezed microwave fields with fluctuations below the shot-noise level can be produced with fully feasible parameters from state-of-the-art cavity electromagnonics and magnomechanics experiments. We adopt the following parameters [24, 25, 28, 29]: $\omega_a/2\pi = 10$ GHz, $\omega_b/2\pi = 10$ MHz, $\gamma/2\pi = 10^2$ Hz, $g/2\pi = 10$ MHz, and at a low temperature $T = 20$ mK, while we set $\Delta_m, \Delta_a, \kappa_m, \kappa_a, G$ as free parameters to optimize the squeezing. We note that both the cavity and magnon decay rates can be controlled on demand. The former can be realized by adjusting the external coupling to the cavity ($\kappa_1/2\pi$ up to 34.4 MHz was achieved [45]), and the latter by changing the loop antenna's position relative to the YIG sphere, allowing to vary $\kappa_m/2\pi$ from 2 MHz to 25 MHz [40]. The choice of the magnon-drive detuning $\widetilde{\Delta}_m$ is nontrival. We adopt a small red detuning $0 < \widetilde{\Delta}_m < \omega_b$, enlightened by the cavity optomechanical experiments [9–12] that produced squeezed light, and the fact that the magnetostrictive interaction is a radiation pressure-like interaction. A red detuning $\widetilde{\Delta}_m > 0$ also helps to stabilize the system as it effectively activates the magnomechanical cooling of the phonon mode [31].

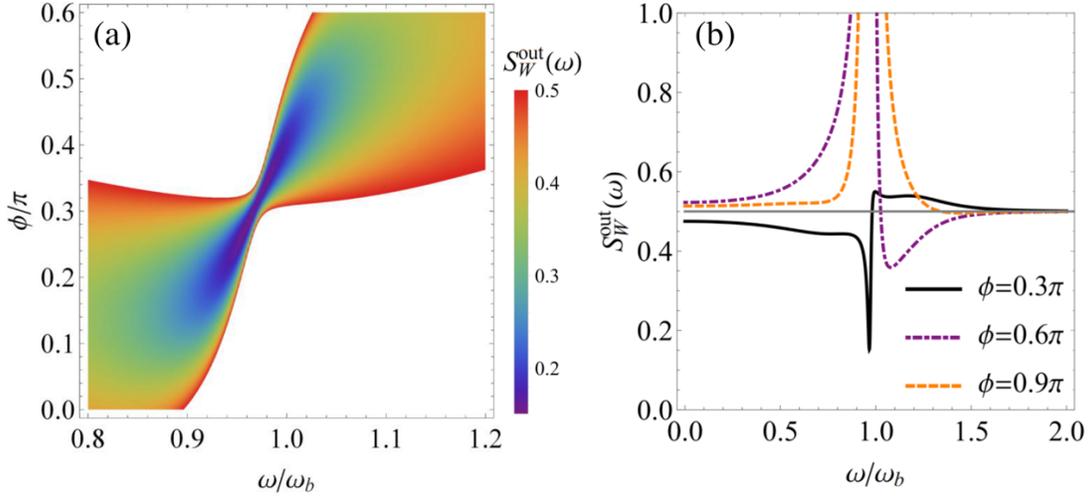

**Figure 2.** (a) Noise spectral density $S_W^{\text{out}}(\omega)$ of the cavity output field versus frequency $\omega$ and phase $\phi$. The blank area denotes $S_W^{\text{out}}(\omega) > 0.5$, i.e., above shot-noise (vacuum fluctuation). (b) $S_W^{\text{out}}(\omega)$ at three specific phases: $\phi = 0.3\pi$, $0.6\pi$, and $0.9\pi$. The gray horizontal line denotes vacuum fluctuation.

The squeezing is first created by the correlation between the amplitude and phase of the magnon mode induced by magnetostriction, which can be termed as ponderomotive-like squeezing. Note that, it is known that the cavity-magnon linear coupling $g(am^\dagger + a^\dagger m)$ solely does not produce any squeezing. Therefore, the squeezing in the system can only be generated by the nonlinear magnomechanical coupling. The magnon squeezing is then transferred to the microwave cavity field via their beamsplitter (state-swap) interaction, as sketched in Fig. 1(b). A cavity-magnon strong coupling $\kappa_m, \kappa_a < g$ has been proved to efficiently accomplish this quantum state-transfer process [46, 47]. Therefore, we consider the parameter regime $2\pi * 2$ MHz $\leq \kappa_m, \kappa_a \leq g = 2\pi * 10$ MHz in our simulation and avoid using a much larger coupling $g$ for the stability reason (though strong coupling $g \gg \omega_b$ has been realized, e.g., in Refs. [28, 29]). The stability of the system is guaranteed by the negative eigenvalues (real parts) of the drift matrix (see the online supplementary material), and all the results presented in this work satisfy this condition, and are thus in the steady state.

In Fig. 2(a), we show the NSD of the cavity output field $S_W^{\text{out}}(\omega)$ versus frequency $\omega$ and phase $\phi$. A remarkable stationary squeezing appears around mechanical frequency $\omega \simeq \omega_b$ and phase $\phi \simeq 0.3\pi$, corresponding to a minimum $S_W^{\text{out}} \simeq 0.15$ (5.2 dB below vacuum fluctuation). In Fig. 2(b), we show $S_W^{\text{out}}(\omega)$ for three phases, $\phi = 0.3\pi$, $0.6\pi$, and $0.9\pi$. While $\phi = 0.3\pi$ gives a maximal squeezing, $\phi = 0.9\pi$ yields a negligible squeezing with $S_W^{\text{out}}(\omega) > 0.495$ in the whole range. Note that an optimal phase $\phi_{\text{opt}}$ can in principle be derived if the NSD of the amplitude and phase quadratures $S_X^{\text{out}}(\omega)$ and $S_Y^{\text{out}}(\omega)$, and their symmetrized correlation spectrum $S_{XY}^{\text{out}}(\omega)$ are known [12]. However, in our complex tripartite system, the analytical expressions of these three NSD are too lengthy, forcing us to optimize $\phi$ numerically. Nevertheless, this optimal phase can be easily found by varying the phase of the local oscillator in the homodyne detection. We have used $\widetilde{\Delta}_m = 3\Delta_a = 0.3\omega_b$, $\kappa_a = 5\kappa_m = \omega_b$ ($\kappa_1 = 0.9\omega_b$, $\kappa_2 = 0.1\omega_b$), $G_0/2\pi = 0.1$ Hz, and drive power $P = 100$ mW (power up to ~ 300 mW via a loop antenna was used in [38]). This corresponds to a drive magnetic field $B \simeq 1.3 \times 10^{-4}$ T, Rabi frequency $\Omega/2\pi \simeq 3.8 \times 10^{14}$ Hz, magnon amplitude $|\langle m \rangle| \simeq 1.9 \times 10^7$, and thus effective coupling $|G| \simeq 0.19\omega_b$, for a 125-μm-radius YIG sphere employed in [24]. Therefore, $|\langle m \rangle|^2 \simeq 3.5 \times 10^{14} \ll 2Ns \simeq 1.7 \times 10^{17}$, satisfying the condition of low-lying excitations. The strong magnon drive may bring about unwanted magnon Kerr nonlinearity into the system [38, 39]. The maximal Kerr coefficient $K/2\pi \simeq 6.4$ nHz for a 125-μm-radius sphere, and $K|\langle m \rangle|^3 \ll \Omega$ must hold in order to keep the Kerr effect negligible [31]. The parameters used in Fig. 2 yield $K|\langle m \rangle|^3/\Omega \simeq 0.11$, which implies that the Kerr nonlinearity can be neglected and our theory model is complete.

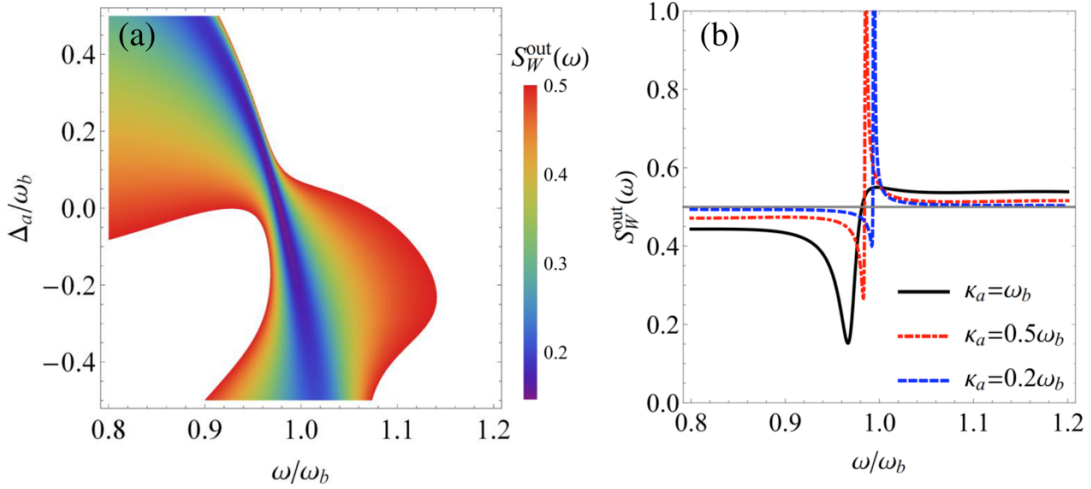

**Figure 3.** (a) $S_W^{\text{out}}(\omega)$ of the output field versus frequency $\omega$ and cavity detuning $\Delta_a$. The blank area denotes $S_W^{\text{out}}(\omega) > 0.5$. (b) $S_W^{\text{out}}(\omega)$ for three cavity decay rates: $\kappa_a = 0.2\omega_b$, $0.5\omega_b$, and $\omega_b$. Other parameters are the same as in Fig. 2, but for $\phi = 0.3\pi$.

In Fig. 3, we further optimize the NSD of the output field $S_W^{\text{out}}(\omega)$ over two cavity parameters, i.e., cavity-drive detuning $\Delta_a$ (Fig. 3(a)) and cavity decay rate $\kappa_a$ (Fig. 3(b)), at an optimal phase $\phi \simeq 0.3\pi$ (under the parameters of Fig. 2). It shows that a small detuning $|\Delta_a| < \omega_b$ and $\kappa_m < \kappa_a$, corresponding to nearly resonant cavity and magnon modes, are optimal for realizing state transfer from magnons to cavity photons [46, 47], This is generally true after numerically exploring the whole period of $\phi$.

**DISCUSSIONS**

We note that, although Figs. 2 and 3 are obtained under a low temperature $T = 20$ mK, the squeezing is still present at $T > 0.5$ K (see Fig. 4(a)), which is much higher than the operation temperatures (tens of mK) of a JPA [19–22] typically adopted for producing squeezed microwave fields. A higher mechanical $Q$ factor allows us to obtain squeezing at an even higher temperature: the squeezing exists at up to 0.75 K for $Q = 10^6$ (taking $\gamma/2\pi = 10$ Hz). In Figs. 2, 3, and 4(a), we have used a coupling rate $G_0/2\pi = 0.1$ Hz, larger than the demonstrated 10 mHz in [24] for a 125-μm-radius YIG sphere, but smaller than the measured 0.2 Hz using the ferromagnetic material CoFeB [42]. Although a (much) larger $G_0$ can be achieved, as explained, by using a YIG film, we show that even with such a weak coupling $G_0/2\pi \simeq 10$ mHz, the squeezing can still be achieved at $T > 0.1$ K, as shown in Fig. 4(b). This leads to an effective coupling $|G| \simeq 0.019\omega_b = 2\pi \times 190$ kHz under the parameters of Fig. 2, which is about 6 times stronger than the coupling rate 30 kHz achieved in [24], partially because we adopt a more efficient direct pump for magnons.

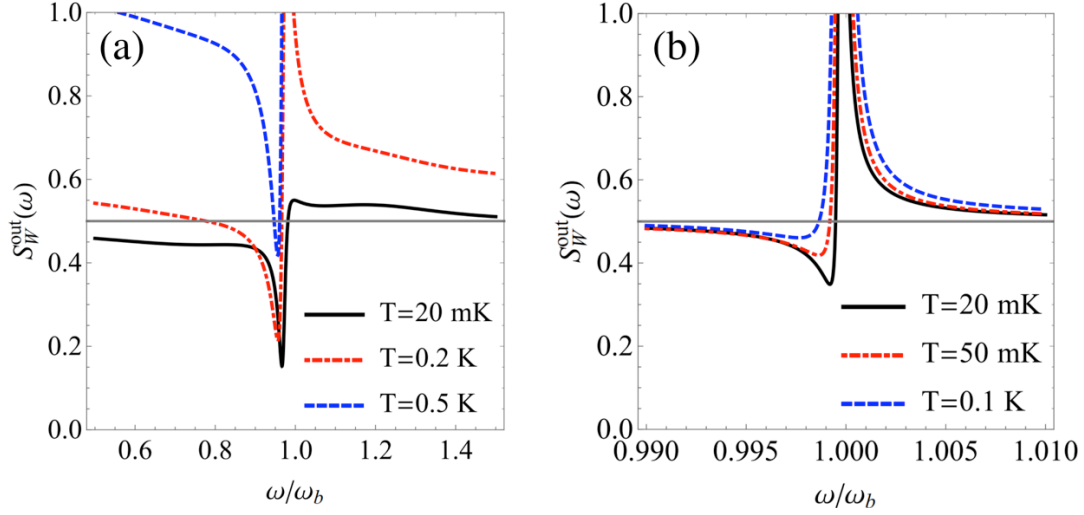

**Figure 4.** $S_W^{\text{out}}(\omega)$ of the output field for (a) $G_0/2\pi = 0.1$ Hz and at various temperatures: $T = 20$ mK, 0.2 K, and 0.5 K, and for (b) $G_0/2\pi = 10$ mHz and at temperatures: $T = 20$ mK, 50 mK, and 0.1 K. In each case, phase $\phi$ is optimized for obtaining $S_W^{\text{out}}(\omega)$. Other parameters are the same as in Fig. 2.

To obtain a more pronounced squeezing, major efforts should be devoted to experimentally improve the magnomechanical cooperativity $\mathcal{C} = |G|^2/(\kappa_m \gamma)$, as such kind of ponderomotive-like squeezing increases with the cooperativity and the squeezing can be potentially large [32, 33, 48]. Although the mechanical damping rate $\gamma$ can be small, the magnon damping rate $\kappa_m$ is typically hard to be below $2\pi \times 1$ MHz due to the intrinsic loss of YIG. More likely, efforts should be devoted to improve the effective coupling $G$, i.e., the bare coupling $G_0$ and the pump power $P$. A large power, however, can cause the system to be unstable and a significant magnon Kerr effect. We note that the Kerr coefficient $K$ can be set to be nearly zero by adjusting the angle between the bias magnetic field and the crystal axes of YIG [49]. Then the pump power is only restricted by the stability condition. Under the parameters of Fig. 2, the maximum power allowed by the stability condition is $\sim 370$ mW, which yields a maximum squeezing of 5.6 dB. This means that the improvement by simply increasing the power is actually limited. We'd like to point out that our model and its full theoretical predictions are generally valid for any magnon mode and mechanical mode that have a radiation pressure-like interaction described by (1), but are not limited to a specific ferromagnet of a certain shape. The analogous successful demonstrations in diverse optomechanical systems [9–12] clearly show such potential possibilities.

Lastly, it would be beneficial to compare our approach with other efficient approaches, e.g., [13], used for producing squeezed microwave fields. The microwave squeezing in [13] was achieved by applying the reservoir engineering to a microwave optomechanical system. The squeezing is termed as dissipative squeezing, distinguished from ponderomotive squeezing [9–12]. Although the dissipative squeezing can be substantial in the good cavity limit $\kappa/\Omega_m \ll 1$ (optimally works at $\kappa/\Omega_m \to 0$, with cavity linewidth $\kappa$ and mechanical frequency $\Omega_m$) [48], the squeezing degrades significantly when $\kappa/\Omega_m \ll 1$ is not fulfilled, which is still experimentally challenging for many optomechanical systems, e.g., those used in [9–13]. In contrast, the optimal condition for our approach, $\kappa_m < \kappa_a \leq g$, can be easily satisfied in current cavity magnonic systems owing to the excellent properties of YIG (low damping rate and high spin density) and a highly tunable cavity decay rate.

## CONCLUSIONS

We provide a new and efficient approach for producing squeezed vacuum states of microwave fields based on a cavity magnomechanical system and by exploiting the nonlinear magnetostrictive interaction in a ferrimagnet. We show that squeezed microwave fields can be obtained with fully feasible parameters from recent experiments in cavity electromagnonics and magnomechanics. We provide optimal parameter regimes for achieving a substantial and stationary squeezing robust against environmental temperature. Our work indicates that cavity magnomechanical systems could act as a novel and promising platform for producing nonclassical states of electromagnetic fields, and may find a wide range of applications in quantum metrology and quantum information processing, such as in improving the sensitivity of a variety of measurements including, e.g., position measurement [50] and magnetic resonance spectroscopy [51], and in producing microwave entangled states [52], etc.

## SUPPLEMENTARY DATA

Supplementary data are available at *NSR* online.

## ACKNOWLEDGEMENTS

The authors thank D. Vitali for reading the manuscript and providing helpful suggestions.

## FUNDING

This work was supported by Zhejiang Province Program for Science and Technology (Grant No. 2020C01019), the National Natural Science Foundation of China (Grants Nos. U1801661, 11934010, 12174329), and the Fundamental Research Funds for the Central Universities (No. 2021FZZX001-02).

## AUTHOR CONTRIBUTIONS

J.L. conceived the idea and did the calculation, and discussed with Y.P.W. on experimental realization. J.L. wrote the manuscript and all authors commented on and revised the manuscript.

*Conflict of interest statement*. None declared.